\documentclass[useAMS,usenatbib]{mn2e}

\usepackage{graphicx}
\usepackage{epsfig}
\usepackage{subfigure}

\newcommand{\be}{\begin{equation}}
\newcommand{\ee}{\end{equation}}

\def \aap{AAP}
\def \apjl{ApJ}
\def \apj{ApJ}
\def \apjs{ApJS}
\def \araa{AAP}

\def \mnras{MNRAS.}

\def \nat{Nature}
\def\msun{{\,M_\odot}}

\def\lesssim{\lower.5ex\hbox{$\; \buildrel < \over \sim \;$}}
\def\gtrsim{\lower.5ex\hbox{$\; \buildrel > \over \sim \;$}}

\title[Galactic Centre Wind]{Gamma-Rays and the Far-Infrared--Radio Continuum Correlation Reveal a Powerful Galactic Centre Wind}
\author[Roland M. Crocker, David Jones, Felix Aharonian, et al.]
{R.~M.~Crocker$^{1}$\thanks{E-mail:Roland.Crocker@mpi-hd.mpg.de}\thanks{Marie Curie Fellow},
D. I. Jones$^{1}$, 
F. Aharonian$^{2,1}$, 
C. J. Law$^{3}$,
F. Melia$^{4}$,  
and J. Ott$^{5}$\\
$^{1}$Max-Planck-Institut f{\" u}r Kernphsik, P.O. Box 103980 Heidelberg, Germany\\
$^{2}$Dublin Institute for Advanced Studies, 31 Fitzwilliam Place, Dublin 2, Ireland\\
$^{3}$Radio Astronomy Lab, University of California, Berkeley, CA 94720, U.S.A.\\
$^{4}$Physics Department, The Applied Math Program, and Steward Observatory, The University of Arizona, Tucson, AZ 85721, U.S.A.\\
$^{5}$National Radio Astronomy Observatory, P.O. Box O 1003, Lopezville Rd, Socorro, NM 87801, USA}
\begin{document}


\date{Accepted XXX. Received XXX; in original form XXX}

\pagerange{\pageref{firstpage}--\pageref{lastpage}} \pubyear{2002}

\maketitle

\label{firstpage}

\begin{abstract}
We consider the thermal and non-thermal emission from the inner 200 pc of the Galaxy. 
The radiation from this almost star-burst-like region is ultimately driven dominantly by on-going massive star formation.
We show that this region's radio continuum (RC) emission is in relative deficit with respect to the expectation afforded by the Far-infrared--Radio Continuum Correlation (FRC).
Likewise we show that the region's $\gamma$-ray emission falls short of that expected given its star formation and resultant supernova rates.
These facts are compellingly explained by positing that a powerful (400-1200 km/s) wind is launched from the region. 
This wind probably plays a number of important roles including advecting positrons into the Galactic bulge thus explaining the observed $\sim$kpc extension of the 511 keV positron annihilation signal around the GC.
We also show that the large-scale GC  magnetic field falls in the range $\sim$100-300 $\mu$G and that -- in the time they remain in the region -- GC cosmic rays do not penetrate into the region's densest molecular material.
\end{abstract}



\section{Introduction}

The extreme ISM conditions in the central $\sim$ 200 pc of the Galaxy render the region more akin to a star-bursting system  \citep[e.g.,][]{Launhardt2002} than to 
almost any region in the Galactic disk.
The similarities 
include: 
i) a high areal star-formation and (consequent) supernova rates; 
ii) a flatish overall radio spectrum within the star-forming region \citep[cf.][]{Niklas1997,Thompson2006}; 
iii) a region surrounding the star-forming nucleus of bright but diffuse, non-thermal radio emission; 
iv) the existence of diffuse $\gamma$-ray emission also apparently associated with star formation \citep[cf.][on NGC 253 and M82]{Abdo2009,Acero2009,Acciari2009}; and
v) a rather strong magnetic field \citep[$>50 \ \mu$G;][]{Crocker2010}.
Here we argue for another similarity: a strong outflow with a speed 400-1200 km/s (comparable to the escape speed) and energetically consistent with being driven by current star-formation \citep{Veilleux2005,Strickland2009}.

Massive, young stars are copious producers of UV and optical light which is reprocessed into IR emission by the dust of the stars' natal molecular envelopes \citep{Devereux1990}.
On the other hand, 
cosmic ray (CR) electrons and ions -- 
ultimately powered by supernovae \citep[e.g.,][]{Hillas2005} -- produce their own (non-thermal) radiative signatures. 
These include $\sim$GHz radio continuum (RC)  synchrotron emission and
inverse Compton (IC) and bremsstrahlung emission at $\gamma$-ray wavelengths by CR electrons and
$\gamma$-rays from neutral meson decay following hadronic collisions between 
CR
ions and gas.

\vspace{-0.1cm}

Given the connection of these radiative processes back to massive ($M_\star > 8 \msun$) star formation \citep{Voelk1989},  
one might expect that they be globally correlated. 
Such is observed  \citep{Dickey1984,deJong1985,Helou1985}: an extremely tight (dispersion of $\sim 0.26$ dex; Yun et al. 2001) FIR-RC correlation (FRC) is found \citep[e.g.,][]{Condon1992} to hold over five orders of magnitude in RC luminosity \citep{Yun2001}, and both globally and at sub-galactic scales \citep{Hughes2006,Tabatabaei2007}.
Likewise, one might also expect  \citep{Thompson2006,Thompson2007}
a global scaling between FIR and $\gamma$-ray production  (`F$\gamma$S').
As we show below, however -- and in interesting contrast to star-bursting systems \citep{Thompson2006}  -- 
the GC does not fall on these scaling relations: we detect far less non-thermal emission than expected given the region's star-formation rate.
This deficit is ultimately explained by a large-scale, powerful outflow from the region.

\section{Correlations and Scalings}
\label{sctn_Correlations}

The H.E.S.S. Imaging Air-Cherenkov $\gamma$-ray Telescope has detected hard-spectrum, diffuse $\sim$TeV $\gamma$-ray emission surrounding the GC over the region defined  by $| l | < 0.8^\circ$ and $| b | < 0.3^\circ$ with an intensity of $1.4  \times 10^{-20}$ cm$^{-2}$ eV$^{-1}$ s$^{-1}$ sr$^{-1}$ at 1 TeV (with the point TeV source coincident with Sgr A$^*$ subtracted).
Only dimmer diffuse TeV emission is detected outside this (hereinafter) `HESS field'.

Of note, is that the spectral index, $\gamma$, of the GC diffuse $\sim$TeV emission, where $F_\gamma \propto E_\gamma^{-\gamma}$, is $2.3 \pm 0.07_\textrm{\tiny{stat}} \pm 0.20_\textrm{\tiny{sys}}$, significantly harder than the spectral index of the CR ion population threading the Galactic disk 
and the  diffuse $\gamma$-ray emission it generates.
Disk CRs experience 
energy-dependent confinement and their steady-state distribution is, therefore, steepened from the 
injection distribution into the softer $\sim E^{-2.75}$ spectrum observed at earth \citep[see, e.g.,][]{Aharonian2006}. 
The GC TeV $\gamma$-ray spectral index (and that inferred for the parent CR ions) is close to that inferred for the {\it injection} spectrum of Galactic disk CRs, itself within the reasonable range of
$\sim$ 2.1-2.2 expected \citep{Hillas2005} for 1st-order Fermi acceleration at astrophysical shocks. 

Empirically the 1.4 GHz RC  (spectral) luminosity and the total IR  luminosity ($L_{TIR}[8-1000] \ \mu m$; \citealt{Calzetti2000}) are connected as \citep{Yun2001,Thompson2007}
\be
\nu L_\nu(1.4 \ \textrm{GHz}) \simeq 1.1 \times 10^{-6}  \ L_{TIR}
\ee
with a scatter of $\sim 0.26$ dex. 
On the basis of IRAS data \citep{Launhardt2002}
the $L_{TIR}$ of the HESS field is $1.6 \times 10^{42}$ erg/s, implying \citep{Kennicutt1998} a SFR of 0.08  $\msun$/yr for the HESS field.
In useful units, the 1.4 GHz RC luminosity \citep[]{Reich1990} of the HESS field  is $1.7 \times 10^{35}$ erg/s[\footnote{We have removed the contribution from synchrotron emission from relativistic electrons in the Galactic plane but out of the GC: see Crocker, Jones, Aharonian et al. 2010, {\it to be submitted}, henceforth Paper II.}], {\it $\sim$1.0 dex or $\sim4 \sigma$ short of the expectation from the
FRC}.

\citet{Thompson2007} 
use the empirically-established connection between the SFR and the total infrared luminosity
to relate the power, injected by supernovae into CRs, to $L_{TIR}$ and hence to
{\it predict} that the TIR and $\gamma$-ray emission from luminous star-forming galaxies should scale as
\be
\nu L_\nu(\textrm{GeV}) \simeq  2.0 \times 10^{-5}  \ \eta_{0.10} \ L_{TIR}
\label{eqn_GeVFIR}
\ee
where the proton spectrum is assumed $\propto E_p^{-2}$ up to $E_p^\textrm{\tiny{max}} \simeq 10^{15}$ eV and we have renormalized the equation of  \citet{Thompson2007} assuming $\eta_{0.10}$ 10\% of the $10^{51}$ ergs per supernova goes into relativistic ions.
This relation assumes that the region under consideration is calorimetric to CR ions.

On the basis of the results presented by \citet{Meurer2009}, Fermi observes a luminosity of $\sim 3 \times 10^{36}$ erg/s for $E_\gamma >$ GeV for emission from the central $1^\circ \times 1^\circ$ field,
only $\sim$10\% of that expectated from the FIR emission.
The Fermi observations are, however, substantially polluted by line-of-sight and point source emission \citep[including from a source coincident with Sgr A*:][]{Chernyakova2010} so they only constitute an upper limit to the true diffuse $\gamma$-ray emission from the region.

We can consider the HESS data by scaling eq.~\ref{eqn_GeVFIR} from $L_\gamma(E_\gamma >$ GeV) to $L_\gamma(E_\gamma >$ TeV). 
For the TeV spectral index of $\sim 2.3$ and assuming an hadronic origin to the TeV $\gamma$-rays, $L_\gamma(E_\gamma >$ TeV) $\simeq 0.2 \ L_\gamma(E_\gamma >$ GeV). 
The TeV luminosity we infer for the HESS field of 1.2$\times 10^{35}$ erg/s (integrating to 100 TeV)
{\it is only $\sim$2\% of the prediction from the suitably-scaled version of eq.~\ref{eqn_GeVFIR}.}

Thus the FRC fails badly in the case of the HESS field: far less RC than expected is detected given its FIR output. 
Likewise, the $\gamma$-ray luminosity of the region is significantly in deficit given the region's FIR output (and implied star-formation rate).
There are three potential explanations of these discrepancies:

Firstly, a RC deficit  could arise if a starburst event occurred more recently ($\lesssim 10^7$ years) than the lifetime of the massive stars whose supernova remnants accelerate the CR electrons which generate synchrotron emission. 
Although we expect some stochastic variation in the GC's overall
SFR, we find, however, that the current SFR is close to the long-term ($\gtrsim 10^7$) average value \citep[cf.][]{Serabyn1996,Figer2004}.
A strong piece of evidence for this is that a number of other handles on the GC supernova rate we describe in Paper II that are sensitive to long-term average values of this quantity -- through, e.g., studies of the region's pulsar population \citep{Lazio2008} -- are consistent with the supernova rate implied by the  current SFR as traced by FIR, viz. 0.04/century in the HESS field.

Secondly, it may be that  GC SNRs are intrinsically low-efficiency \citep[cf.][]{Erlykin2007} CR accelerators  \citep[plausible because of their -- {\it on average} -- dense environs:][]{Fatuzzo2005}.
However,  the  detailed numerical modelling set out in Paper II shows
that GC supernovae do, indeed, accelerate CRs with typical \citep[e.g.,][]{Hillas2005} efficiency: about 10\% of the total $10^{51}$ erg mechanical energy per supernova goes into non-thermal particles.
Given the above rate, this implies that supernovae 
inject $ \sim 10^{39}$ erg/s into the GC CR population \citep[cf.][]{Crocker2010b}.

Lastly, given the half-height of the region is only $\sim 40$ pc, a reasonable reaction to the breakdown of the FRC is that it is unsurprising; many studies  \citep{Murgia2005} find a break-down in the correlation at $\sim$ kpc, often proposed to be due to electron transport.  
On the other hand, studies, e.g., of the Large Magellanic Cloud \citep{Hughes2006}, the Scd galaxy M33 \citep{Tabatabaei2007}, and within the Milky Way \citep{Zhang2010} reveal a tight connection between RC and FIR emission down to scales $\lesssim$ 50 pc.

A potential
fourth explanation of why the HESS field falls off the FRC is that power fed into non-thermal electrons is `lost' to ionization and bremsstrahlung and/or inverse Compton emission \citep{Thompson2006,Thompson2007} rather than synchrotron emission (plausible because of the GC's dense gas and radiation environment).
Given, however, the HESS field also falls short of the F$\gamma$S, this explanation is, at least, seriously incomplete.

In summary here, it seems that {\it CR transport out of the HESS field is by far the most plausible explanation for why it falls off the global scalings discussed}; below we show that the transport mechanism is a wind.

\section{Prior Evidence for an Outflow from the GC}
\label{sectn_outflow}

There is multi-wavelength
evidence in support of the existence of GC outflow.
Recent infra-red observations show that the GC's massive stellar clusters are blowing a bubble into their environment \citep[e.g.,][]{Bally2010}.
\citet{Keeney2006} and \citet{Zech2008} have found evidence for high-velocity gas consistent with a GC outflow or fountain in UV absorption features towards, respectively, two AGN and a GC globular cluster.
The region's spectacular non-thermal radio filaments \citep{Yusef-Zadeh1987} 
may be due to a fast outflow \citep[e.g.,][]{Shore1999}.
RC evidence of an outflow
was found in 10 GHz radio continuum emission by \citet{Sofue1984} in the form of a $\sim1^\circ$ (or $\sim 140$ pc) tall and diameter $<130$ pc shell of emission rising north of the Galactic plane called the Galactic Centre lobe (GCL).
RC emission from the lobe's eastern part has HI absorption that clearly puts it in the GC region \citep{Lasenby1989} and its ionized gas has a high metalicity \citep{Law2009}.
Filamentary structures coincident with the radio have been discovered at mid-infrared wavelengths \citep{Bland-Hawthorn2003} and 
the structure interpreted as evidence for a previous episode of either starburst \citep{Bland-Hawthorn2003} or nuclear activity \citep{Melia2001}.
\citet{Law2010} has found that the formation of the GCL is consistent with currently-observed pressures and rates of
star-formation in the central few $\times 10$ pc  of the Galaxy.
Finally, \citet{Law2010} determined the $\sim$GHz spectral index of the GCL steepens with increasing distance (both north and south) of the Galactic plane. 
This constitutes strong evidence for synchrotron ageing of a CR electron population transported out of the plane.
Thus, a natural interpretation is that the GCL's RC emission is due to 
CR electrons advected from the inner GC (essentially the HESS region) on a wind \citep[cf.][on, e.g., NGC 253 and M82]{Zirakashvili2006,Heesen2009}.

\begin{figure}
 \epsfig{file=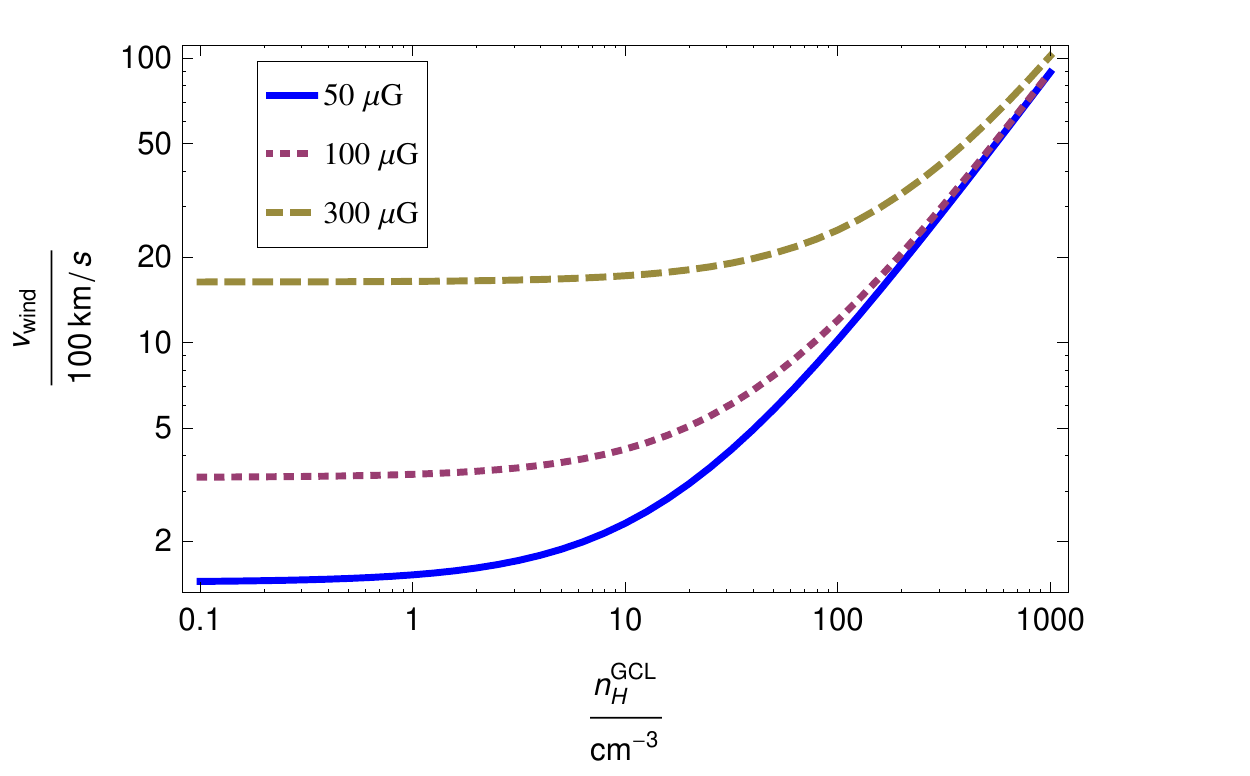,width=\columnwidth}
\caption{Lower bound on the wind speed required for electrons advected out of HESS field to synchrotron-illuminate the entire extent of the GC lobe within their loss times given by ionization, bremsstrahlung, synchrotron, and IC emission for environmental parameters of $B$ and $n_H$ and an interstellar radiation field energy density $U_{ISRF} \simeq 20$ eV cm$^{-3}$.
We infer from  \citet{Ferriere2007} that the volumetric average $n_H$ in the GCL is $\sim 10$ cm$^{-3}$.
The strict lower limit to the GCL magnetic field at 50 $\mu$G \citep[and probable value 100 $\mu$G:][]{Crocker2010} imply a conservative lower limit to the GC outflow speed of $>$ 150 km/s (and probably $\gtrsim 300$ km/s). 
}
\label{plotRqrdWindSpeed}
\end{figure}

This interpretation requires that 
\begin{itemize}
\item
The spectrum of the electrons leaving the HESS region (as given by Eq.\ref {solutionSmpl}) must match the spectrum {\it at injection} required for the GCL electrons with spectral index 2.0 -- 2.4 \citep{Crocker2010}.
This will be well-satisfied if an energy-independent transport process like a wind removes CR electrons -- accelerated into an in-situ $\sim E^{-2}$ distribution -- from the inner GC.
\item
The power in electrons leaving the HESS region must be enough to support the GCL electron population, viz. $(3-10) \times 10^{37}$  erg/s \citep{Crocker2010}.
This is well satisfied given the SN rate in the HESS region.
\item The time to transport electrons over the extent of the GCL must be less than the loss time over the same scale.
This implies a wind speed of strictly $>$ 150  km/s and probably $\gtrsim$ 300  km/s: see fig.~\ref{plotRqrdWindSpeed}.
\end{itemize}

\section{Non-Thermal Hints of an Outflow from the GC}
\label{sctn_Hints}

An important consideration is why the GC CR ion population is so hard in comparison to the diffusion-steepened, local population.
There are three reasonable interpretations of this:
i) the system is out of steady state with  less time having passed since the CR injection event than required for diffusion steepening \citep[cf.~the interpretation adopted by][that a single CR-injection event $\sim10^4$ year ago at the GC explains the observed diminution in the $\gamma$-ray to molecular column ratio beyond $|l| \sim 1^\circ$]{Aharonian2006}; {\it or} there is a smallest relevant timescale defined by an energy-independent ii) {\it loss} or iii) {\it escape} process.

We argue here that iii) is preferred by all the evidence. 
We can dismiss ii) on the basis of our results above  which show the system falls far short of being a calorimeter for protons.
A number of factors also tell against i):
firstly, as argued, other evidence indicates that the system is close to its steady state;
secondly, 
the  spectral index of the $\sim$TeV emission is a constant $\sim 2.3$ over the HESS region (within errors) presenting, therefore, 
no evidence of diffusion hardening at the leading edge of a (putative) diffusion sphere; and 
lastly, these spectral considerations apply also to the relativistic electron population: the hard radio spectrum of the region, 
$\alpha \lesssim 0.54$ (for $S_\nu \propto \nu^{-\alpha}$ and radio data 1.4--10 GHz: see Paper II), 
requires that the synchrotron-emitting electron population is also very hard, $\sim E_e^{-2.1}$. 
Given the rather short loss times associated with synchrotron and IC emission in the GC environment, this hard electron spectrum constitutes independent evidence for {\it rather quick and energy-independent CR transport} \citep[cf.][]{Lisenfeld2000}.

Consider then the region's non-thermal particle population which,
in steady state, approximates to: 
\be
\label{solutionSmpl}
n_x\!(E_x) \simeq \frac{\tau_\textrm{\tiny{loss}}\!(E_x) \tau_\textrm{\tiny{esc}}}{\tau_\textrm{\tiny{loss}}\!(E_x)+(\gamma-1)\tau_\textrm{\tiny{esc}}} {\dot Q}_x\!(E_x)
\ee
where ${\dot Q}_e(E_e)$ denotes 
the injection rate of particles  of type $x \in \{e,p\}$; we account for both escape and energy loss 
over $\tau_\textrm{\tiny{esc}}$ and $\tau_\textrm{\tiny{loss}}$ with the escape time assumed to be energy-independent; and
$\gamma$ is the spectral index of the (assumed) power-law (in momentum) proton or electron spectrum at injection.

Turning now to the CR ion population (henceforth protons for simplicity), we have already seen that we only detect $\sim$2\% of the TeV $\gamma$-ray flux expected in the calorimetric limit. 
Given, then, that $pp$ collisions are by far the dominant energy loss process for high-energy CR protons, this deficit implies that there is significant escape of accelerated ions (with accompanying adiabatic losses) -- i.e., the system is quite far from calorimetric.
We define $R_\textrm{\tiny{TeV}} \equiv  L_\textrm{\tiny{TeV}}^\textrm{\tiny{obs}}/L_\textrm{\tiny{TeV}}^\textrm{\tiny{thick}} \simeq 10^{-2}$ (uncertain by a factor $\sim$2) as the ratio of the observed flux of TeV $\gamma$-ray emission to the expected  in the calorimetric limit
\citep[cf. fractions  $\sim 0.01$ and $\sim 0.05$  for the Galactic disk and NGC 253;][]{Acero2009}. 
From Eq.\ref{solutionSmpl} and accounting for adiabatic losses with timescale $\tau^p_\textrm{\tiny{adbtc}} = 3 \tau^p_\textrm{\tiny{esc}}$:
\be
R_\textrm{\tiny{TeV}} \simeq 10^{-2}
\sim \frac{3 \tau_\textrm{\tiny{esc}}^{p}}{3 \tau_\textrm{\tiny{esc}}^{p} + 4 \tau_{pp}}\ .
\label{eqn_hadronic}
\ee

By analogy with the hadronic case, we define $R_\textrm{\tiny{radio}} \equiv  L_\textrm{\tiny{synch}}^\textrm{\tiny{obs}}/L_\textrm{\tiny{synch}}^\textrm{\tiny{thick}}  \simeq 10^{-1}$ (again uncertain by a factor $\sim$2).
Given the very flat radio spectral index, this deficit is potentially explained as a result of electron energy loss into bremsstrahlung, adiabatic deceleration or advective escape.
Eq.\ref{solutionSmpl} then gives
\be
R_\textrm{\tiny{radio}} \simeq 0.1 \sim 
\frac{\tau_\textrm{\tiny{esc}}^{e} ( \tau_\textrm{\tiny{brems}} + 3 \tau^e_\textrm{\tiny{esc}})}{ \tau_\textrm{\tiny{synch}}(\tau_\textrm{\tiny{brems}} + 4\tau_\textrm{\tiny{esc}}^{e} )}
 \ .
\label{eqn_leptonic}
\ee

\begin{figure}
 \epsfig{file=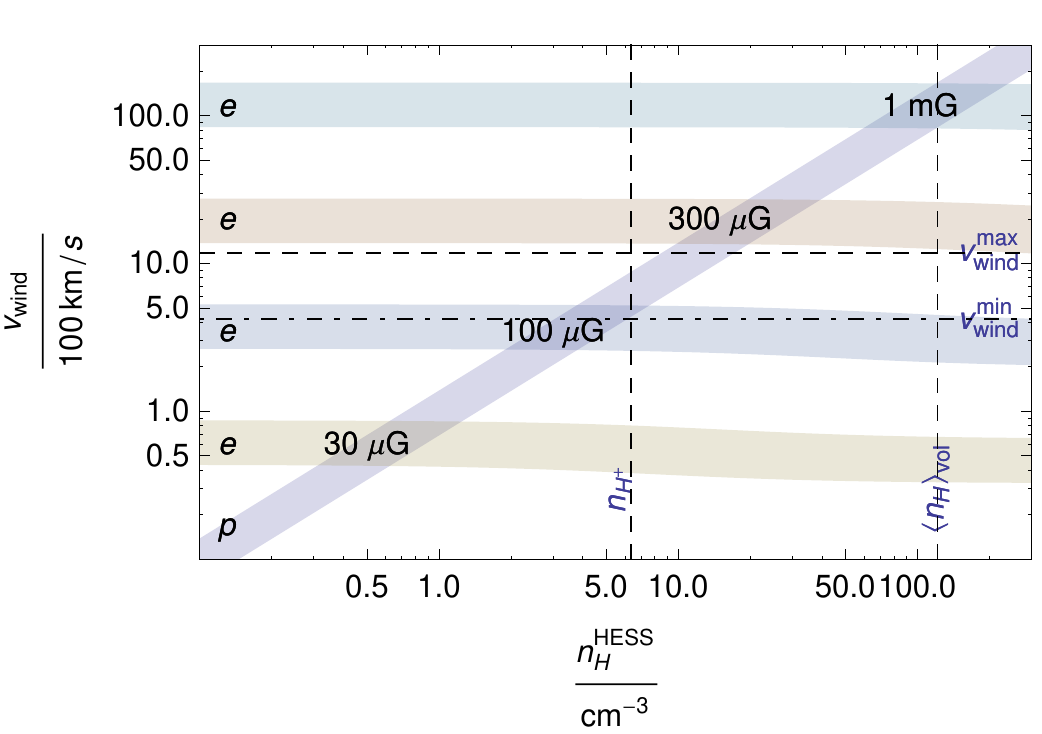,width=\columnwidth}
\caption{Outflow speed inferred given the departures from calorimetry for both protons (`$p$') and electrons (`$e$'):  $R_\textrm{\tiny{TeV}} = 0.01 $ and $R_\textrm{\tiny{radio}} \simeq 0.1$ as described in the text (the width of 
the bands reflects the uncertainty of $\sim$2 in both $R_\textrm{\tiny{TeV}}$ and $R_\textrm{\tiny{radio}}$).
Protons cool via their hadronic collisions with ambient gas (hence the linear dependence between wind speed and gas density, $n_H$)
and adiabatic deceleration. 
In addition to bremssrahlung (and ionization),  electrons also cool via synchrotron (so the magnetic field enters as a parameter) and IC emission and adiabatic deceleration. 
As the wind escape time is the same for both electrons and protons, {\it the intersection of  the electron and proton bands describes a valid gas density and wind velocity for the HESS environment for each magnetic field sampled}. 
The horizontal dashed line shows the approximate maximum allowed wind speed ($\sim1200$ km/s) balancing the total power assumed injected into the system by supernovae and massive stars ($1.4 \times 10^{40}$ erg/s) with the kinetic power advected by the wind plasma at its asymptotic velocity (assuming 100\% thermalization efficiency). 
The horizontal dot-dashed line shows the approximate minimum plausible wind speed ($\sim400$ km/s) for thermalization efficiency of 10\%.
}
\label{plotvWind}
\end{figure}

Now, given the foregoing, particle escape is both energy-independent and the same for CR electrons and protons ($\tau_\textrm{\tiny{esc}}^{e} \equiv \tau_\textrm{\tiny{esc}}^{p} =$ const) as would be expected for a wind. 
This means that eq.~\ref{eqn_hadronic} and \ref{eqn_leptonic}
yield a combined constraint on the required velocity of the outflow responsible for particle removal: see fig.~\ref{plotvWind}. 

Also shown in fig.~\ref{plotvWind} are minimum and maximum values for the speed of the star-formation-driven
`super-wind' expected on the basis of observations of the nuclei of external, star-forming galaxies 
and the GC's high areal SFR \citep{Strickland2009}.
 The asymptotic speed of such a wind 
 scales as
 $v_{\tiny{wind}} \sim \sqrt{2 \ \eta \ \dot{E}/\dot{M}}$ where $0 < \eta < 1$ is the thermalization efficiency, typically ranging between 0.1 for relatively quiescent star formation and almost 1 for star-bursts \citep{Strickland2009}.
 Adopting  $\eta_{\tiny{min}} \equiv 0.1$, $\eta_{\tiny{max}} \equiv 1.0$,  $\dot{E} = 1.4 \times 10^{40}$ erg/s and 0.025 $\msun$/year (see Paper II), we find $v_{\tiny{wind}}^{min} \simeq 400$ km/s and $v_{\tiny{wind}}^{max} \simeq 1200$ km/s.
 
\begin{figure}
 \epsfig{file=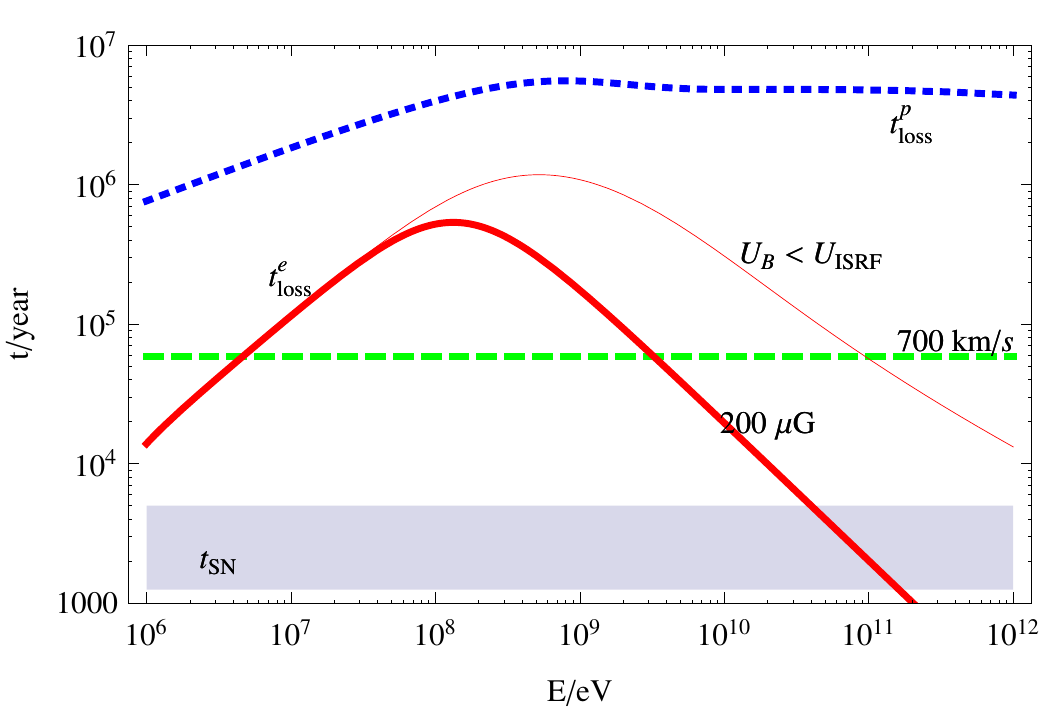,width=\columnwidth}
\caption{HESS region timescales for central parameter values suggested by 
our analysis, viz.  $n_H = 10$ cm$^{-3}$, and $v_\textrm{\tiny{wind}} = 700$ km/s
with
i) (horizontal solid band) the inverse of the supernova rate;
ii) (dashed horizontal line) particle escape  with energy-independent velocity of 700 km/s;
iii) (solid red lines) electron cooling for (thick) $B = 2 \times 10^{-4}$ G and (thin) the limiting case of vanishing magnetic field (IC cooling dominant at high energy);
iv) (blue dotted line) proton cooling.
Calorimetry generically requires $t_{loss} < t_{esc}$.
}
\label{plotTimescales}
\end{figure}

Putting some of these considerations in a different form, we expect a TeV luminosity from the HESS region which satisfies
$
L_\gamma (E_\gamma > \textrm{TeV}) \sim 1/3  \ U_\textrm{\tiny{CR}}(E_p > 10 \ \textrm{TeV})/\tau_{pp} \ V \leq L_\gamma^\textrm{\tiny{obs}} (E_\gamma > \textrm{TeV}) \equiv 1.2 \times 10^{35} \ \textrm{erg/s}$ 
where 
$U_\textrm{\tiny{CR}}(E_p > 10 \ \textrm{TeV}) \sim 1/20 \times 1.4 \times 10^{39}$ 
erg/s 
$\times \ d/v_{\tiny{wind}}/V$ 
is the energy density in CR protons sufficiently energetic to generate TeV $\gamma$-rays,
$d \simeq 40$ pc and $V \simeq 10^{62}$ cm$^{-3}$ for the HESS region, and $n_H$ is the {\it effective} gas density the protons sample. 
This implies $n_H \lesssim 6$ cm$^{-3} \ (v_{\tiny{wind}}/1200$ km/s), cf. the volumetric average gas density through the HESS region $\sim$120 cm$^{-3}$ summing over all phases and  $\sim$6 cm$^{-3}$ including only plasma phases.
Likewise, the total gas mass the protons sample satisfies $M_{\tiny{gas}} \lesssim 5 \times 10^5  \msun  \ (v_{\tiny{wind}}/1200$ km/s) which is much less than the $\sim 10^7 \msun$ of gas in the region.
In order that the region's protons {\it not} sample all the molecular gas in the region they should be removed in a time shorter than the convection time into the 
dense regions of the molecular clouds: $t_{\tiny{wind}} \equiv d/v_{\tiny{wind}} < t_{\tiny{cloud}} \sim 10$ pc/30 km/s \citep[adopting 30 km/s as a typical internal velocity dispersion for the region's giant molecular clouds, conservatively, of radius $\sim$10 pc: e.g.][]{Morris1996} which also implies a lower limit: $v_{\tiny{wind}} \gtrsim 130$ km/s.
 Typical timescales are plotted in fig.~\ref{plotTimescales}.

\section{Discussion and Conclusions}
\label{sctn_Discussion}

A clear picture emerges from the considerations above. 
Given the morphological and spectral data on the GC lobe, we can infer that it is illuminated with CR electrons injected in the HESS region carried from the plane on an outflow with a speed 150--1000 km/s.
The spectral data on the HESS region itself imply that most CR electrons and protons accelerated in situ
are advected from the region; electrons  lose only $O[10 \ \%]$  of their power to synchrotron emission in the HESS region while protons lose only $O[1 \ \%]$  of their power to $pp$ collisions on ambient gas in the same region.
Self-consistently and given our understanding of outflows from external, star-forming galaxies, the same star-formation and subsequent supernova processes that drive the thermal and non-thermal radiation from the HESS region will also drive an outflow with a speed 400--1200 km/s.
This implies that the magnetic field in the HESS field lies in the range 100-300 $\mu$G and the effective gas density encountered by the CRs is in the range 3--20 cm$^{-3}$.
The latter is much less than the
volumetric average $n_H$ over the HESS region suggesting that even super-TeV CRs do not `sample' all $H_2$ before escaping the region.

We suspect that the outflow we identify plays many important roles \citep[see Paper II and][]{Crocker2010b} including advecting positrons into the Galactic bulge \citep[thereby explaining the $\sim$kpc extension of the 511 keV annihilation radiation:][]{Weidenspointner2008}, carrying CR ions accelerated by GC supernovae out to very large heights ($\sim$ 10 kpc) thereby explaining the WMAP `haze' and Fermi 'bubbles' \citep{Finkbeiner2004,Dobler2009,Su2010,Crocker2010b}, and generally keeping the energy density of the non-thermal components of the GC ISM in check \citep{Breitschwerdt2002}.

\vspace{-0.73cm}
\section{Acknowledgements}

The authors gratefully acknowledge correspondence with  Rainer Beck,  Joss Bland-Hawthorn, Valenti Bosch-Ramon,  Sabrina Casanova, Roger Clay, John Dickey, Ron Ekers, Katia Ferri{\` e}re, Stanislav Kel'ner,  Mitya Khangulyan, Jasmina Lazendic-Galloway, Mark Morris,   Giovanni Natale, Emma de O\~na Wilhelmi,  Ray Protheroe, Brian Reville, Frank Rieger, Ary Rodr{\'{\i}}guez-Gonz{\'a}lez, Gavin Rowell, and  Andrew Taylor. RMC particularly thanks  Heinz V{\"o}lk for many enlightening discussions. The authors are indebted to the referee, Professor John Bally, for an expeditious and insightful review.

\label{lastpage}


\vspace{-0.7cm}


\begin{thebibliography}{}

\bibitem[Fermi LAT Collaboration(2009)]{Abdo2009} Fermi LAT Collaboration
2009, arXiv:0911.5327 

\bibitem[HESS Collaboration(2009)]{Acero2009} The 
HESS Collaboration: F.~Acero 2009, arXiv:0909.4651 

\bibitem[VERITAS collaboration(2009)]{Acciari2009} Acciari, V.~A., et al.\ 
2009, arXiv:0911.0873 

\bibitem[Aharonian et al.(2006)]{Aharonian2006}
{Aharonian}, F.~A. et~al. 2006, Nature, 439, 695

\bibitem[Bally et al.(2010)]{Bally2010} Bally, J., et al.\ 2010, 
\apj, 721, 137

\bibitem[Bland-Hawthorn 
\& Cohen(2003)]{Bland-Hawthorn2003} Bland-Hawthorn, J., \& Cohen, M.\ 2003, \apj, 582, 246 

\bibitem[Breitschwerdt et 
al.(2002)]{Breitschwerdt2002} Breitschwerdt, D., Dogiel, V.~A., V{\" o}lk, H.~J.\ 2002, \aap, 385, 216 

\bibitem[Calzetti et al.(2000)]{Calzetti2000} Calzetti, D. et al., 
2000, \apj, 533, 682 

\bibitem[Chernyakova et al.(2010)]{Chernyakova2010} Chernyakova, M., 
Malyshev, D., Aharonian, F.~A., Crocker, R.~M., 
\& Jones, D.~I.\ 2010, arXiv:1009.2630 

\bibitem[Condon(1992)]{Condon1992} Condon, J.~J.\ 1992, \araa, 30, 575 

\bibitem[Crocker et al.(2010)]{Crocker2010} Crocker, R.~M. et al., 2010
\nat, 463, 65 

\bibitem[Crocker \& Aharonian(2010)]{Crocker2010b} Crocker, R.~M. \& Aharonian, F. 2010,  arXiv:1008.2658

\bibitem[de Jong et 
al.(1985)]{deJong1985} de Jong, T. et al., 
1985, \aap, 147, L6 

\bibitem[Devereux 
\& Young(1990)]{Devereux1990} Devereux, N.~A., \& Young, J.~S.\ 1990, \apjl, 350, L25 

\bibitem[Dickey 
\& Salpeter(1984)]{Dickey1984} Dickey, J.~M., \& Salpeter, E.~E.\ 1984, \apj, 284, 461 

\bibitem[Dobler et al.(2010)]{Dobler2009} Dobler, G. et al.,
2010, \apj, 717, 825 

\bibitem[Erlykin 
\& Wolfendale(2007)]{Erlykin2007} Erlykin, A.~D., \& Wolfendale, A.~W.\ 2007, JPGNP, 34, 1813 

\bibitem[Fatuzzo 
\& Melia(2005)]{Fatuzzo2005} Fatuzzo, M., \& Melia, F.\ 2005, \apj, 630, 321 

\bibitem[Ferri{\`e}re et~al.(2007)]{Ferriere2007} Ferri{\`e}re, K., Gillard, W., \& Jean, P.\ 2007, \aap, 467, 611 

\bibitem[Figer et~al.(2004)]{Figer2004} Figer, D., et al. 2004,  \apjl, 601, 319-339 (2004)

\bibitem[Finkbeiner et~al.(2004)]{Finkbeiner2004} Finkbeiner, D.~P.  et al. 2004, \apj, 617, 350-359

\bibitem[Heesen et 
al.(2009)]{Heesen2009} Heesen, V., et al., 
2009, \aap, 494, 563 

\bibitem[Helou et al.(1985)]{Helou1985} Helou, G., Soifer, B.~T., 
\& Rowan-Robinson, M.\ 1985, \apjl, 298, L7 

\bibitem[Hillas(2005)]{Hillas2005} Hillas, A.~M.\ 2005, JPGNP, 31, 95 

\bibitem[Hughes et al.(2006)]{Hughes2006} Hughes, A. et al.,
\ 2006, \mnras, 370, 363 

\bibitem[Keeney et al.(2006)]{Keeney2006} Keeney, B.~A. et al., 
2006, \apj, 646, 951 

\bibitem[Kennicutt(1998)]{Kennicutt1998} Kennicutt, R.~C., Jr.\ 1998, 
\apj, 498, 541 

\bibitem[LaRosa et~al.(2005)]{LaRosa2005} LaRosa, T.~N. et al.,
2005, \apjl, 626, L23 

\bibitem[Lasenby et al.(1989)]{Lasenby1989} Lasenby, J., Lasenby, 
A.~N., \& Yusef-Zadeh, F.\ 1989, \apj, 343, 177 

\bibitem[Launhardt et~al.(2002)]{Launhardt2002} Launhardt, R., Zylka, R., \& Mezger, P.~G.\ 2002, \aap, 384, 112 

\bibitem[Law et al.(2008)]{Law2008} Law, C.~J. et al.,
2008, \apjs, 177, 255 

\bibitem[Law et al.(2009)]{Law2009} Law, C.~J., Backer, D., 
Yusef-Zadeh, F., \& Maddalena, R.\ 2009, \apj, 695, 1070 

\bibitem[Law(2010)]{Law2010} Law, C.~J.\ 2010, \apj, 708, 474 

\bibitem[Lazio 
\& Cordes(2008)]{Lazio2008} Lazio, T.~J.~W., \& Cordes, J.~M.\ 2008, \apjs, 174, 481 

\bibitem[Lisenfeld
\& V{\" o}lk(2000)]{Lisenfeld2000} Lisenfeld, U. \& V{\" o}lk, H.\ 2000, A\&A 354, 423

\bibitem[Melia 
\& Falcke(2001)]{Melia2001} Melia, F., \& Falcke, H.\ 2001, \araa, 39, 309 

\bibitem[Meurer (2009)]{Meurer2009} Meurer, C.\ 2009, `First constraints on Dark Matter...', TeVPA conference 2009 (SLAC)

\bibitem[Morris \& Serabyn(1996)]{Morris1996} Morris, M., \& 
Serabyn, E.\ 1996, ARA\&A, 34, 645 

\bibitem[Murgia et 
al.(2005)]{Murgia2005} Murgia, M. et al.,
2005, \aap, 437, 389 

\bibitem[Niklas et 
al.(1997)]{Niklas1997} Niklas, S., Klein, U., \& Wielebinski, R.\ 1997, \aap, 322, 19 

\bibitem[Reich et~al.(1990)]{Reich1990} Reich, W., Reich, P., Fuerst, E., 1990, AAPS.,83, 539-568

\bibitem[Reynolds et al.(2008)]{Reynolds2008} Reynolds, S.~P. et al., 
2008, \apjl, 680, L41 

\bibitem[Serabyn \& Morris(1996)]{Serabyn1996} Serabyn, E., \& Morris, M. 1996, \nat, 382, 602

\bibitem[Shore 
\& LaRosa(1999)]{Shore1999} Shore, S.~N., \& LaRosa, T.~N.\ 1999, \apj, 521, 587 

\bibitem[Sofue 
\& Handa(1984)]{Sofue1984} Sofue, Y., \& Handa, T.\ 1984, \nat, 310, 568 

\bibitem[Strickland 
\& Heckman(2009)]{Strickland2009} Strickland, D.~K., \& Heckman, T.~M.\ 2009, \apj, 697, 2030 

\bibitem[Su et al.(2010)]{Su2010} Su, M., Slatyer, T.~R., 
\& Finkbeiner, D.~P.\ 2010, arXiv:1005.5480 

\bibitem[Tabatabaei et 
al.(2007)]{Tabatabaei2007} Tabatabaei, F.~S., et al.\ 2007, \aap, 466, 509 

\bibitem[Thompson et~al.(2006)]{Thompson2006} Thompson, T.~A. et al., 
2006, \apj, 645, 186 

\bibitem[Thompson et al.(2007)]{Thompson2007} Thompson, T.~A., 
Quataert, E., \& Waxman, E.\ 2007, \apj, 654, 219 

\bibitem[Veilleux et 
al.(2005)]{Veilleux2005} Veilleux, S., Cecil, G., \& Bland-Hawthorn, J.\ 2005, \araa, 43, 769 

\bibitem[V{\" o}lk(1989)]{Voelk1989} V{\" o}lk, H.~J.\ 1989, \aap, 218, 67 

\bibitem[Weidenspointner et al.(2008)]{Weidenspointner2008} Weidenspointner, G. et al. 2008, Nature, 451, 159

\bibitem[Yun et~al.(2001)]{Yun2001} Yun, M.~S., Reddy, N.~A., 
\& Condon, J.~J.\ 2001, \apj, 554, 803

\bibitem[Yusef-Zadeh et~al.(1987)]{Yusef-Zadeh1987} 
Yusef-Zadeh, F. \& Morris, M.\ 1987, AJ 94, 1178 

\bibitem[Zech et al.(2008)]{Zech2008} Zech, W.~F., Lehner, N., 
Howk, J.~C., Dixon, W.~V.~D., \& Brown, T.~M.\ 2008, \apj, 679, 460 

\bibitem[Zhang et al.(2010)] {Zhang2010} Zhang, J., et al. \ 2010, P.A.S.A, 27, 340 

\bibitem[Zirakashvili \& V{\" o}lk(2006)]{Zirakashvili2006} Zirakashvili, V.~N., V{\" o}lk, H.~J.\ 2006, \apj, 636, 140 


\end{thebibliography}
\end{document}